# HERMES Pathfinder & SpIRIT: a progress report

**F. Fiore[a]*, M.Trenti[m], Y. Evangelista[a], R. Campana[a], G. Baroni[a], F. Ceraudo[a], M. Citossi[a], G. Della Casa[a], G. Dilillo[a], M. Feroci[a], M. Fiorini[a], G. Ghirlanda[a], C. Labanti[a], G. La Rosa[a], E.J. Marchesini[a], G. Morgante[a], L. Nava[a], P. Nogara[a], A. Nuti[a], M.Perri[a], F. Russo[a], G. Sottile[a],
M. Lavagna[b], A. Colagrossi[b], S. Silvestrini[b], M. Quirino[b], M. Bechini[b], L. Bianchi[b], A. Brandonisio[b], F. De Cecio[b]. A. Dottori[b], I. Troisi[b], G. Bertuccio[b], F. Mele[b],
B. Negri[c], R. Bertacin[c], C. Grappasonni[c], R. Piazzolla[c], S. Pirrotta[c], S. Puccetti[c], M. Rinaldi[c], A. Tiberia[c],
L. Burderi[d], A. Sanna[d], A. Riggio[d], C. Cabras[d], A. Tsvetkova[d],
A. Santangelo[e], A. Guzman[e], P. Hedderman[e], S. Pliego Caballero[e], C. Tenzer[e],
A. Vacchi[f], N. Zampa[fk], R. Crupi[fa],
P. Bellutti[g], E. Demenev[g], F. Ficorella[g], D. Novel[g], G. Pepponi[g], A. Picciotto[g], N. Zorzi[g],
M. Grassi[h], P. Malcovati[h],
T. Di Salvo[i], W. Leone[j,i], S. Trevisan[j,a],
I. Rashevskaya[k], A. Rachevski[k], G. Zampa[k],
T. Chen[l], N. Gao[l], S. Xiong[l], S. Yi[l], S. Zhang[l],
M. Ortiz del Castillo[m], R. Mearns[m], J. McRobbie[m], A. Chapman[m], M. Thomas[m], A. Woods[m], J. Morgan[m], S. Barraclough[m],
N. Werner[n], J. Ripa[n], F. Munz[n], A. Pal[o],
D. Gačnik[p], A. Hudrap[p], D. Selčan[p], G. Molera Calves[q]**

[a] INAF, via del Parco Mellini 84, 00138 Roma, Italy;[b] Politecnico di Milano, Italy; [c]ASI, Italy; [d]University of Cagliari, Italy; [e]University of Tubingen, Germany; [f]University of Udine, Italy; [g]Fondazione Bruno Kessler, Italy; [h]University of Pavia; [i]University of Palermo; [j]University of Trento; [k]INFN; [l]IHEP, China; [m]University of Melbourne; Australia; [n]Masaryk University; [o]Konkoly Observatory, Hungary; [p]Skylabs; [q]School of Natural Sciences, University of Tasmania
* Corresponding Author fabrizio.fiore@inaf.it

**Abstract**

The High Energy Rapid Modular Ensemble of Satellites Pathfinder (HERMES-PF) is an in-orbit demonstration mission consisting of a constellation of six 3U cubesats hosting simple but innovative X-ray/gamma-ray detectors for the monitoring of cosmic high-energy transients. HERMES-PF, funded by the Italian Space Agency (ASI) and by the European Commission (EC) through a Horizon 2020 grant, is expected to be launched at the beginning of 2025. An identical X-ray/gamma-ray detector is hosted by the Australian 6U cubesat Space Industry Responsive Intelligent Thermal nanosatellite (SpIRIT), launched on December 1st 2023. The main objective of HERMES-PF/SpIRIT is to demonstrate that high energy cosmic transients such as Gamma Ray Bursts can be detected efficiently by miniatured hardware and localized using triangulation techniques. The HERMES-PF X-ray/gamma-ray detector is made by 60 GAGG:Ce scintillator crystals and 12 2×5 silicon drift detector (SDD) mosaics, used to detect both the cosmic X-rays directly and the optical photons produced by gamma-ray interactions with the instrument scintillator crystals. This innovative design provides a unique broad band spectral coverage from a few keV to a few MeV. Furthermore, the use of fast GAGG:Ce crystals and of mosaics of small SDD cells allows us to reach an exquisite time resolution better than a microsecond. We will present a progress report on the missions focusing the discussion on the scientific innovation of the project and on the main lessons learned during the project development including: the importance but also the challenges of using distributed architectures to achieve ambitious scientific objectives; the importance of developing critical technologies under science agreements for the realization of innovative, high-performing but low-cost payloads; best use of commercial off the shelf (COTS) technologies in scientific missions. We will finally discuss the prospects of applying these concepts for the creation of an observatory with the ability to cover the entire high-energy sky at all times to search for the high-energy counterparts of gravitational wave events that Advanced LIGO/Virgo/Kagra will find at the end of this decade and the Einstein Telescope during the 2030'.
**Keywords:** (Gamma Ray Bursts, CubeSat, distributed architectures, constellation)

**Acronyms/Abbreviations**

High Energy Rapid Modular Ensemble of Satellites (HERMES), Space Industry Responsive Intelligent

Thermal nanosatellite (SpIRIT), Agenzia Spaziale Italiana (ASI), European Commission (EC), Silicon drift detectors (SDD), Gadolinium Aluminium Gallium Garnet: Cerium (GAGG:Ce), commercial off the shelf (COTS), gravitational wave (GW), gamma-ray bursts (GRB), gravitational wave event (GWE), Einstein Telscope (ET), payload (P/L), spacecraft (S/C), Fondazione Bruno Kessler (FBK), thermal-vacuum (TVAC), Sun-synchronous orbit (SSO), LTAN, Scientific Operation Center (SOC), Mission Operation Center (MOC), space scientific data center (SSDC), flight model (FM), thermal vacuum (TVAC), Sun synchronous orbit (SSO), Launch and Early Orbit Phase, (LEOP), application specific integrated circuit (ASIC).

## 1. Introduction

The past decade has seen the emergence and then the establishment of a new field of investigation in astrophysics, multimessenger astrophysics. It provides a unique opportunity to shed new light on the physics of compact objects, cosmology, and fundamental physics [1]. Since 2015, more than 100 gravitational wave (GW) signals have been discovered: the majority are binary black holes (BH) and four are binaries hosting at least one neutron star (NS). It was postulated for many years that, in the latter case, accretion onto the newly formed object (a BH or a rapidly spinning NS) could lead to the launch of powerful jets. Jet energy dissipation in internal shocks is responsible for a short (lasting less than a few seconds) Gamma-Ray Burst (GRB) [2,3,4]. The jet interaction with the external medium produces the GRB afterglow. Both can be probed by electromagnetic observations [5]. This scenario was demonstrated by the one and only association of GW170817, produced by the merger of two NS, with the short GRB170817. Two seconds after the end of the GW signal, a short GRB was detected and localized by current monolithic instruments (NASA/Fermi and ESA/INTEGRAL) with an accuracy of a few tens of $deg^2$. This allowed us to search and discover the optical/NIR transient (kilonova) and the GRB afterglow (detected from the radio to the X-ray band). The multi-messenger study of GW/GRB170817 taught us that NS-NS mergers can power relativistic jets and are the sites where the heaviest elements in the universe are produced [5,6,7,8,9]. It provided constraints on the equation of state of nuclear matter, the Hubble constant and the precision of the value of the speed of light.

While GRBs probe the physics of particle acceleration and relativistic astrophysical jets, GWs (encoding the rapid/relativistic motion of compact objects) probe the GRB progenitor's properties, e.g., mass, spin, interior properties, inclination, and distance of the systems. Combining these two probes is the core of current multi-messenger astrophysics. To make progress, we need sensitive, high-energy all-sky monitoring, in order to detect and localize the EM counterparts of GW sources and follow them up rapidly with ground and space-based telescopes.

The lessons learned from the first four observing runs (O1, O2, O3, O4) of current GW interferometers are:

(1) an all-sky monitor with a sufficient sensitivity is needed to detect off-axis jets with intrinsic luminosities down to $10^{47}$ ergs/s. Within this decade, the LIGO/Virgo/Kagra interferometers will reach their target sensitivity, and the searched horizon will reach a few hundred Mpc. In the corresponding volumes, the number of optical and near-infrared transients may be as large as hundreds to thousands, hampering the identification of the electromagnetic counterpart. The limited number (a few) of X-ray transients makes the search much more efficient in this band. Considering sensitivity, field of view and duty cycle of Fermi/GBM, we estimate it should detect 0.5–3 short GRBs associated with GW events during O5 (>2028)[10]. The operation of an efficient X-ray all-sky monitor, with a sensitivity comparable to or even better than that of Fermi/GBM, should increase these numbers by a factor ~3–4. The number of GRB detections associated with third generation gravitational interferometers such as the Einstein Telescope (ET) detection can be of several tens to a few hundred per year [11]. The capability to instantaneously cover the whole sky is mandatory, because the number of events will be reasonably small; therefore, missing simply one event would mean a considerable loss for scientific research.

(2) The capability of determining the position of the high-energy transients with uncertainties smaller than a few degrees. Within the volume defined by this spatial constraint and the distance of the GWE provided by the interferometric measurements, the number of optical transients will be small enough to quickly assess the correct transient to associate with the GWE, thus prompting further follow-ups.

Today, X-ray and gamma-ray monitors dedicated to the search and localization of high-energy transients are mostly monolithic instruments (e.g., NASA Swift/BAT, ESA INTEGRAL/IBIS, INTEGRAL/SPI) or multiple detectors hosted by the same large spacecraft (NASA FERMI/GBM). All have been launched during the first decade of 2000 and are aging. It is unclear if all or some of them could support the multimessenger efforts at the end of this decade and during the 2030s. Distributed architectures of nano-small-sats equipped with state-of-the-art X-ray and gamma-ray detectors offer a unique opportunity to develop a sensitive all-sky monitor in time to harvest the first fruits of multimessenger astrophysics bringing it to maturity [12].

Distributed architectures have been used for GRB science since the beginning of these activities. The Interplanetary network (IPN) provided and still provides localizations by using the delay time of arrival of the transient signal on different detectors. The diverse gamma-ray instrumentation used by the IPN, the poor knowledge of the spacecraft position when outside the Earth's GNSS infrastructure, as well as the difficulty of defining an absolute time with a good precision for all spacecrafts and the delay in communication with remote solar system spacecrafts, significantly limit the ability of the IPN to routinely provide accurate (better than a few degrees) and timely localizations. In other words, systematic errors associated with IPN transient localization are usually much greater than the statistical errors. All of these difficulties could be overcome by a constellation of nano-satellites in LEO hosting similar, if not identical, X-ray and gamma-ray detectors. The GPS and Galileo infrastructures offer the opportunity to constrain LEO satellite positions to within a few tens of meters, and the absolute time within a few tens of nanoseconds, orders of magnitude better than what is possible to achieve outside of the GNSS infrastructure. Adopting identical detectors ensures similar responses to cosmic events, reducing systematic uncertainties.

HERMES-PF and SpIRIT were conceived at the end of the 2010' to prove the feasibility of these concepts, that is (1) to show that *miniaturized instrumentation on board CubeSats can routinely detect GRBs*, and (2) to show that *the goal of obtaining positions with accuracy the order of a degree or less is achievable by a constellation of CubeSats* by studying and limiting systematic effects in the transient position determination using the triangularization technique. HERMES-PF and SpIRIT are, therefore, in orbit demonstrations, preparing for relevant scientific data production. However, they are also pathfinders for what concerns the procurement and system activities. The project additional goals of the projects are: *(3) Demonstrate the COTS applicability to challenging space missions* to contain the costs and the time-to-orbit. HERMES-PF will assess and apply a production life cycle to increase the COTS reliability still limiting the time-to-orbit and the cost. (4) *Contribute to the Space 4.0 goals and expectations* by identifying and standardizing new and innovative approaches to manufacture, assemble and test miniaturized components, allowing proactive small and medium-sized enterprises (SMEs) to access the business. (5) *Enlarge and strengthen the space-distributed architectures and mega-constellations applicability and reliability* by investigating and optimizing the space-ground segment nets design in terms of the number of satellites vs the number of ground stations, constellation architecture and coordination and space segments functionalities, space segment reliability, costs, time to launch, time to operation, scalability. The study will deepen the knowledge on potentials and limitations of large-scale, small-platform, distributed architectures' applicability to challenging space science missions.

## 2. HERMES-PF progress report

HERMES-PF is an in-orbit demonstration consisting of a constellation of six 3U CubeSats hosting simple but innovative X-ray/gamma-ray detectors for the monitoring of cosmic high-energy transients such as GRB and the electromagnetic counterparts of GWEs [13]. It is funded by ASI and the EC through a Horizon 2020 grant and it should be launched by Q1 2025. The HERMES-PF X-ray/gamma-ray detector is described in detail in [14,15]. It is made by 60 GAGG:Ce scintillator crystals and 12 2×5 SDD mosaics, used to detect both cosmic X-rays and optical photons produced by gamma-rays in the scintillator crystals. This innovative design provides a unique broad band spectral coverage from a few keV to a few MeV. Furthermore, the use of fast GAGG:Ce crystals and mosaics of small SDD cells allows reaching an exquisite time resolution of <0.4 microsecond. Evaluations of the expected in orbit background and collecting area are presented in [16], while [17] presents the results of the P/L flight units on ground calibrations. P/L on board S/W has been written and tested by the IAAT team at University of Tubingen and it is presented in [18,19].

The HERMES-PF S/C includes all components usually found in satellites. Its internal configuration consists of three units: the Top Unit hosting the P/L; the Central Unit hosting the batteries, the electrical power supply (EPS) docking board including the power distribution unit (PDU), array conditioning unit (ACU), the attitude determination and control (ADCS ), docking board, an on board computer (OBC) dedicated to the attitude control, the main OBC, the UHF/VHF and S-band transceivers and the magnetorquers; the Bottom Unit hosting the reaction wheels, the UHF/VHF antenna the IMU and the Iridium transceiver. The S-band antennae, GPS antenna and Iridium antenna are mounted on the S/C lateral panels. Two-wings solar panels provide the needed power. Further information about the HERMES SVM and mission planning can be found in [20,21,22,23,24,25].

All six P/L flight units have been integrated, tested and calibrated during 2022 and 2023 at Fondazione Bruno Kessler (FBK) and INAF laboratories. At the time of writing, five complete (P/L and S/C) flight models (FM) were integrated and tested at Politecnico di Milano (PoliMI) laboratories. Four FM have already passed environmental tests (mechanic test and thermal vacuum, TVAC, tests) at PoliMI and Thales Alenia Spazio – Gorgonzola laboratories respectively. The integration and test of all six FMs is expected to be completed by the end of October 2024. Fig. 1 and 2 below show H2 fully integrated and ready for the TVAC tests.

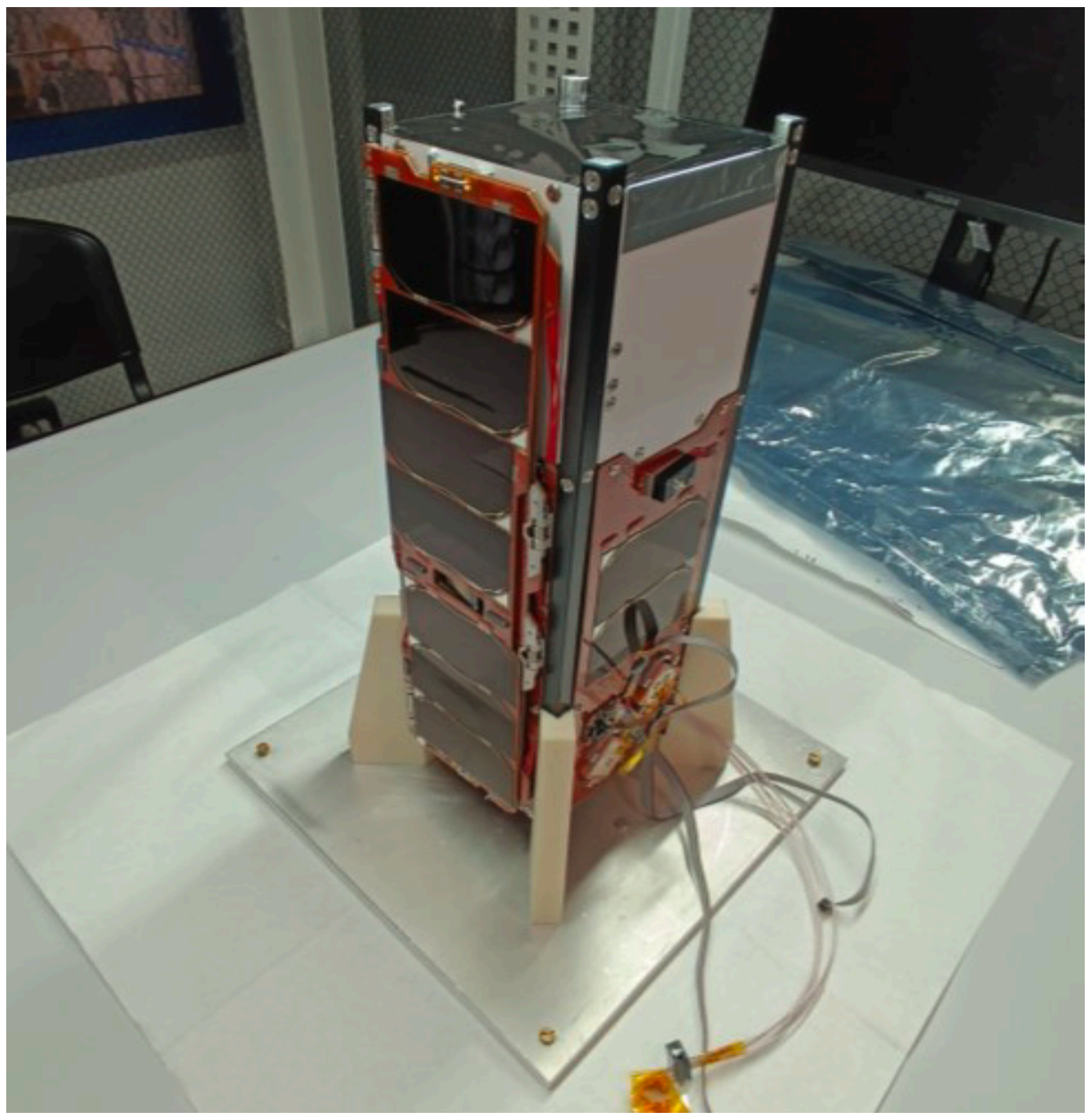

Fig. 1. HERMES-PF H2 FM completely integrated at PoliMI laboratories. The P/L is visible in the top U, a folded solar panel is visible to the left. The S-band antennas are visible to the bottom right,

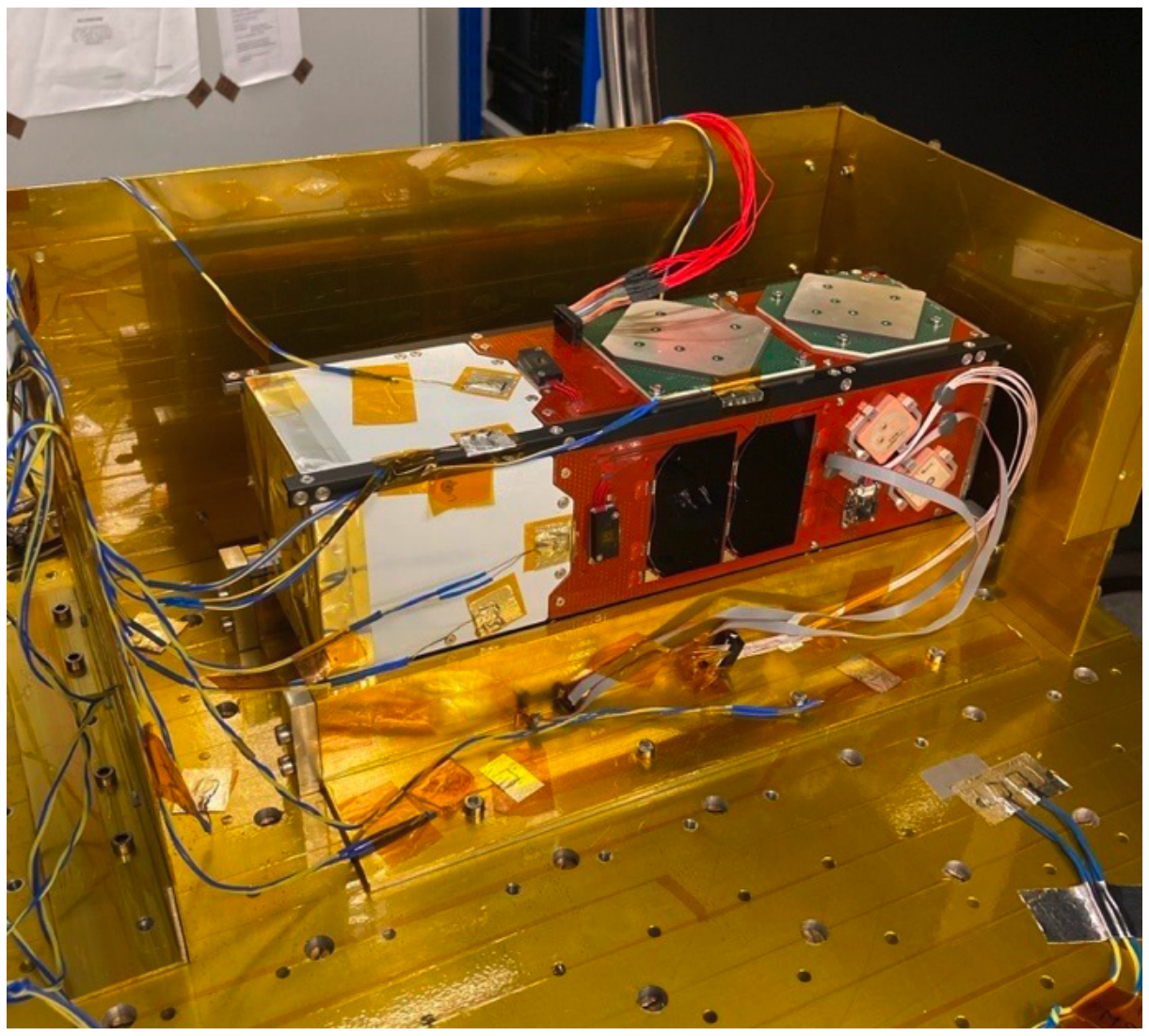

Fig. 2 HERMES-PF H2 FM ready for TVAC environmental tests at Thales Alenia Spazio laboratories in Gorgonzola.

The transient position is obtained by studying the delay time of arrival of the signal to different detectors hosted by nano-satellites on low Earth orbits [26, 27]. The ground system and the Scientific Operation Center is presented by [28,29]. Expected in-flight performances of HERMES-PF are presented in [11].

The launch of HERMES-PF has been procured through an ASI contract to D-Orbit. HERMES-PF will fly on SpaceX Transporter 13 in March 2025. It will be deployed on a polar SSO with a height of 510 km and LTDN at 10:30. The HERMES-PF MOC is managed by Altec through an ASI contract. The SOC is hosted by ASI SSDC. Two dedicated ground stations (both including UHF/VHF and S-band antennas) will support the mission, the first in Spino D'Adda, Italy, and the second in Katherine, Northern Territories, Australia. The first is funded by ASI, the second is funded through the H2020 Infrastructure grant AHEAD2020. Other commercial and university-based ground stations will likely support both the preliminary phase (LEOP, commissioning) and the normal operations.

**3. SpIRIT progress report**

SpIRIT is a 6U CubeSat led and developed by the University of Melbourne (PI Michele Trenti) with funding from the Australian Space Agency (ASA) and in cooperation with the Italian Space Agency and associated entities of the HERMES-PF consortium [30]. SpIRIT hosts a suit of P/Ls managed by a centralized instrument control unit (called Payload Management System - PMS in short). These include one of the X-ray and gamma-ray spectrometers and an S-band transceiver identical to those hosted by HERMES-PF and provided by ASI, an active cooling system based on a Stirling cycle cryocooler (TheMIS [31]), Mercury, a communication subsystem optimized to ensure low-latency short burst data transfer, utilizing the Iridium satellite network [32]; LORIS, visible and IR inspection cameras and an on-board Artificial Intelligence image-processing module [33]. TheMIS is designed to actively cool the HERMES P/L, allowing to operate the SDDs at a low temperature (around –20 °C), to reduce the noise due to radiation damage. Fig. 3 shows the HERMES-PF X-ray and gamma-ray FM before the integration into the SpIRIT satellite.

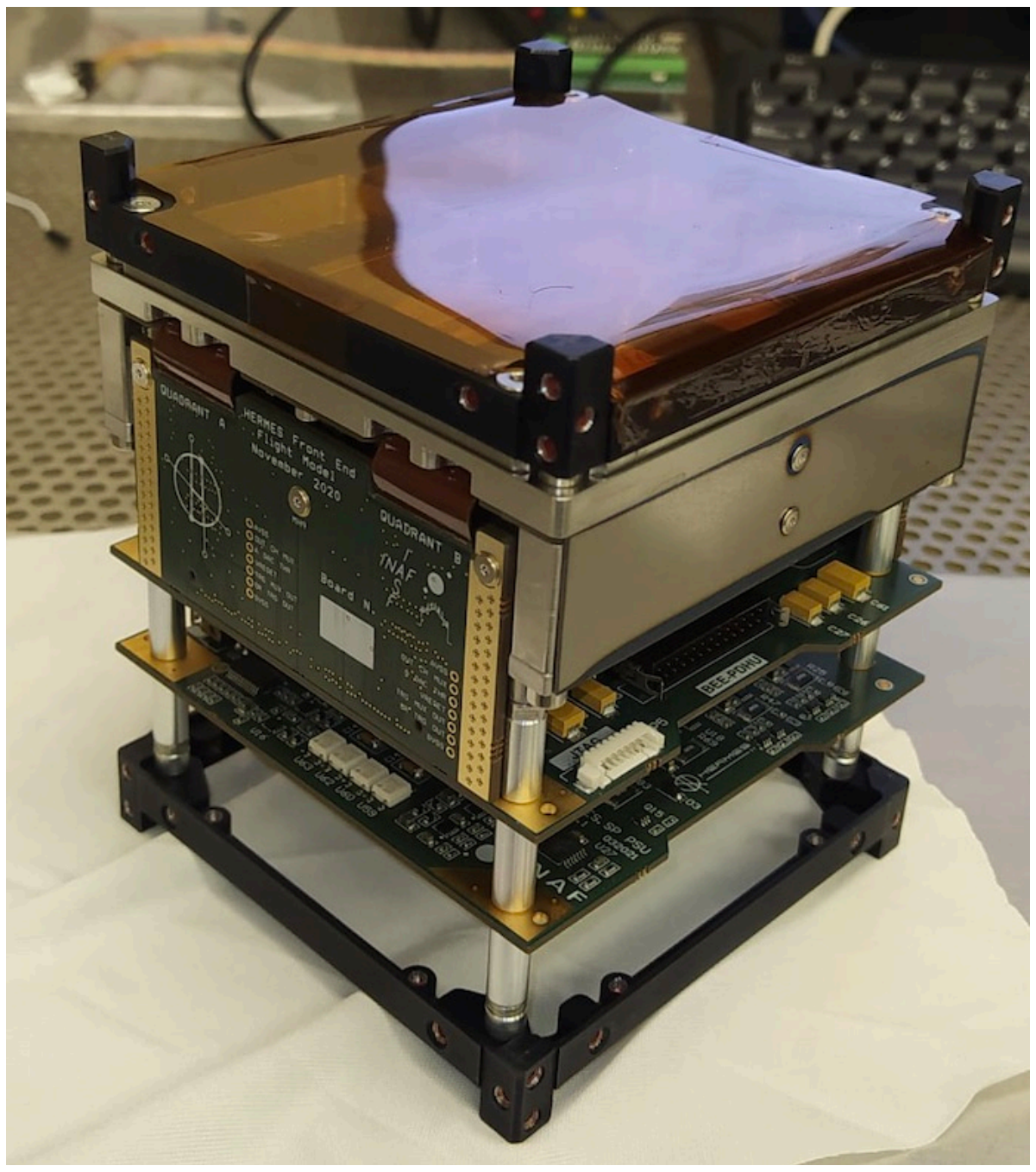

Fig. 3. The P/L FM before integration in the SpIRIT satellite

SpIRIT was successfully launched into a polar SSO (513 km) on December 1st, 2023 from the Vandenberg Space Force Base on a SpaceX Falcon 9. The commissioning of the S/C started soon after the acquisition of communications with the satellite. It is still on going for what regards the S-band and the X-ray/gamma-ray spectrometer. After initial checks of the spacecraft payload control unit and the HERMES-PF payload data management unit, the detectors were activated, achieving "first light" on January 16th, 2024 with the instrument operating in a basic photon counting mode. HERMES-PF successfully entered nominal observation mode on March 27th, 2024, recording scientific data and achieving TRL 9. Details on the commissioning of the HERMES-PF P/L on board SpIRIT can be found in [34]. The P/L so far could be operated for a total of about 100 minutes, primarily without resorting to active cooling due to the desire to simplify operations as much as possible during the satellite commissioning phase. Due to the unavailability of the S-band system data transmission had to resort to the UHF link with the single station in Australia that is allowed to communicate with the satellite. This hampered the transmission of complete temporal and spectral photon data, and limited the data received to housekeeping and scientific ratemeters (light curves in three energy bands, 5–10 keV, 10–100 keV and >100 keV, with 100 ms time bin). Fig. 4 shows a collection of light curves observed by SpIRIT in the three energy bands.

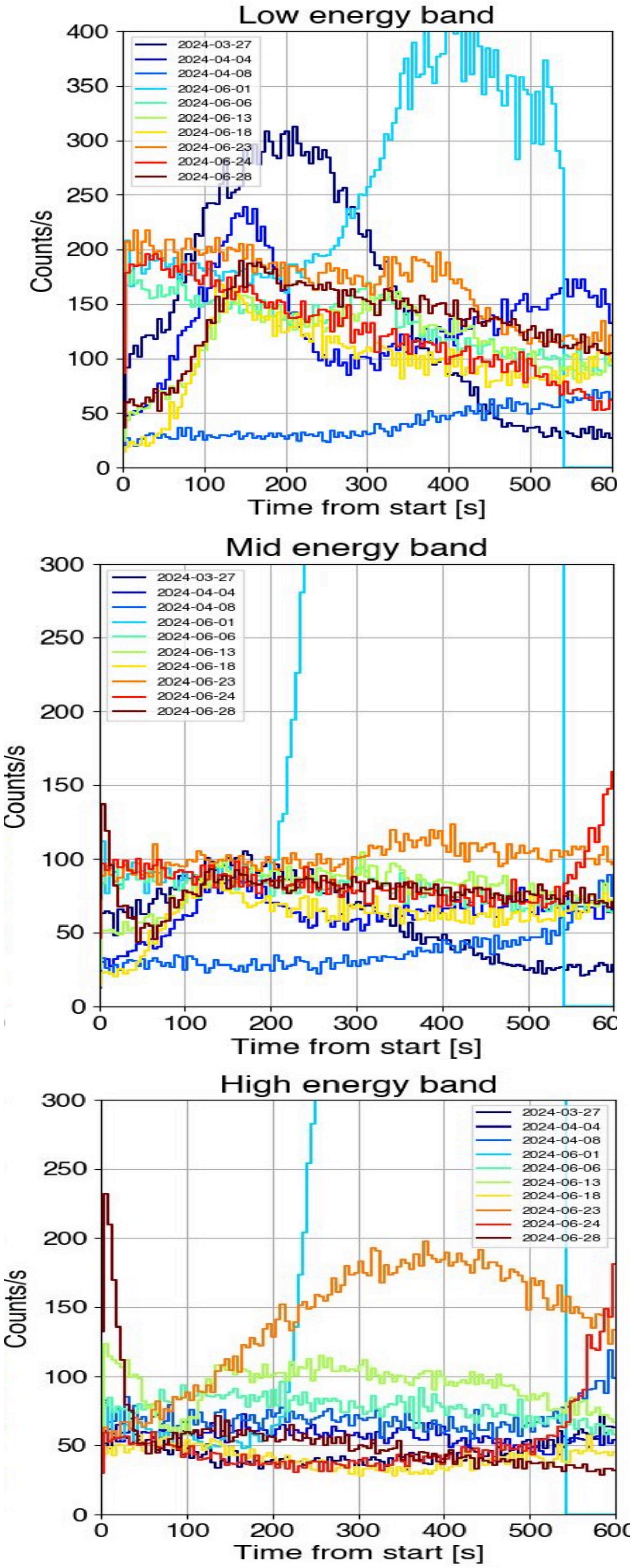

Fig. 4 The light curves of the ten observations performed so far by SpIRIT. The cyan curve refers to an observation where the satellite crossed the North polar electron belt, a region of high particle background.

Despite the limited amount of observing time, these data allowed us to assess that:

a. the payload is functioning correctly;
b. housekeeping data (voltages and currents) are nominal;
c. within the limits of the poor statistics available, photon-by-photon data is consistent with the detector front-end configuration;
d. because of radiation damage and of the relatively high temperatures of most observing runs, the detector noise level is currently high, leading to non-negligible electronic noise counts affecting the ratemeters;
e. there is an issue in the onboard software that produces the ratemeters. Mitigating actions are ongoing;
f. average count rates appear to be roughly consistent with the pre-launch simulated in-orbit background. Specifically, the cosmic X-ray background plus bright Galactic sources dominate the low energy band, and particle background dominates the mid and high energy bands.

The S-band system should come online in the next few weeks for the transmission of scientific data on the ground stations of Katherine and Hobart (managed by University of Tasmania). This will greatly accelerate the P/L operations by eliminating the the data transmission bottleneck. Furthermore, SpIRIT is expected to transition towards full operations of HERMES P/L, which will include active cooling during data acquisition.

## 3. Lesson learned

We discuss here the main lesson learned in the development of the HERMES-PF and SpIRIT projects, which can help conceiving and planning more ambitious projects, such as a complete constellation that can serve as an all-sky monitor for high-energy transients [12,35].

1. Critical technologies for the development of the P/L are those underlying the design and the production of the SDDs, the ASICs to read out and analyse the signal recorded by the SDD, and the thin optical filters needed to avoid optical and UV Sunlight reaching the sensors. We have proceeded with designing and producing critical components through scientific agreements. The SDDs have been designed by the University of Udine, INFN and FBK, and were produced and tested by FBK and INFN. The ASICs were designed by the Department of Electronics, Information and Bioengineering (DEIB) of PoliMI and the Department of Electrical, Computer and Biomedical Engineering of the University of Pavia. ASICs were produced through EUROPRACTICE, a consortium of European research organizations providing universities and research institutes with access to IC prototyping services. Four batches were designed, produced and tested on breadboards from 2018 to 2022, to improve reliability and performance iteratively. Optical filters were designed and produced by IHEP China. They were tested at IHEP and INAF laboratories, confirming a excellent transparency in X-rays and optical UV absorption better that $10^{-10}$. University of Udine, INFN, University of Pavia, PoliMI, FBK and IHEP are all part of the HERMES-PF consortium. The scientific collaboration agreements allowed both full interaction and integration among the teams working on the critical technologies with the scientific team to optimize the design, tailor it to the specific application, and keep costs compatible with a CubeSat project.
2. Several of the COTS components were procured through contracts with SMEs including non-recurring engineering phases to tailor the hardware to the specific application. Key elements of the S/C, the main OBC and the UHF/VHF and S-band systems were provided by SkyLabs, one of the partners of the EC H2020 project. This again allowed close interactions between the science team and the teams working at the design and procurement of the subsystems. Having conducted extensive testing before acceptance of components and subsystems ensured that their integration could proceed quickly and smoothly, limiting risks and loss of time.
3. Although small, the satellites will produce substantial volumes of data. As a reference, each of the HERMES-PF P/L will produce about 0.5-1Gbit/day of data, an amount similar to that produced by BeppoSAX, a large 1.5-ton satellite hosting eight scientific instruments. And, of course, HERMES-PF + SpIRIT are 7 satellites that must work simultaneously. At the rate we can operate the S-band, we would need the order of 3–4 passages per day to download the data acquired by one satellite in one day, that is a total of 18–24 passages per day for six satellites. To cope with these demanding needs two dedicated ground stations were procured and installed, one in Italy and one in Australia. Activities on the ground system (MOC, SOC, and ground stations) where included since the beginning in the EC H2020 HERMES-SP project and entered in the ASI led project at a later stage, through an industrial contract to Altec and a scientific agreement with INAF, PoliMI and University of Cagliari. Handling a constellation of 6 or 7 satellites efficiently is probably the most challenging part of the project from both the technological and scientific points of view. As for the first, the MOC should be able to control the constellation planning and executing about two dozen of passages per day. The SOC should be able to process this large amount of data in real time , and to correlate the data produced by different satellites when they detect a transient. The experience of HERMES-PF/SpIRIT would provide a crucial lesson learned for planning the control and exploitation of future applications of scientific constellations. The ground system must be planned and designed in advance, simultaneously with the flight system.

4. Another very critical element of space missions is the on-board S/W. The P/L on board S/W has been designed and written by the University of Tubingen IAAT team. Requirements were discussed jointly with the INAF team and tests were carried out jointly by the IAAT and INAF teams. This allowed very large flexibility, leading to an efficient iterative development of the S/W, continuously improving performances and fixing bugs. The development and test of the S/W greatly benefited from the early (2020) production of a complete P/L Demonstration Model. The S/C on board S/W was designed by the POLIMI team and developed using an SDK provided by Bright Ascension. The development and test of the S/C S/W greatly benefited from the deployment of engineering models of the most critical S/C elements first and of a "flat-sat" using both flight models and engineering models subsystems at a later time. End to End tests and day-in-a-life tests of the fully integrated FMs, foreseen in the next few weeks, will finally validate both P/L and S/C S/W. The availability of engineering and demonstration models of the hardware as soon as possible is mandatory to limit the risks of S/W failures, bugs or limitations.
5. HERMES-PF and SpIRIT do not use expensive rad-hard electric, electronic and electro-mechanical (EEE) components. HERMES-PF S/Cs use flight-proven components, HERMES-PF P/Ls uses a flight-proven OBC (the PDHU), provided by ISIS-Space. The rest of the P/L's electronic boards have been designed using measures to make them robust in the space environment, such as redundancy and circuits to protect against latch-ups, because they do not use space-qualified components. The most critical components (DC-DC and ADC converters, current sensitive amplifier, LDO regulators, inverters, flip-flop, load switch and other elements) have been validated through radiation tests with a proton beam at INFN TIFPA laboratories in Trento. All P/L electronic boards on board SpIRIT have been working smoothly for more than eight months in the harsh space environment, validating their design and production procedures. The TIFPA proton beam was also used to gain insight into the behavior of both the GAGG:Ce scintillator crystals and of the SDD, when exposed to proton doses representative of typical values encountered on orbit during the whole operative life of up to 2 years [36,37], confirming that GAGG:Ce and SDD can be reliably used in space.
6. The P/Ls have been integrated and tested at INAF and FBK laboratories. The complete FMs, S/C integrated with the P/L, have been integrated and tested at POLIMI laboratories. A university and a scientific institute have acted as integrators without strong involvement of industry (the TVAC tests were executed at Thales Alenia Space Gorgonzola facility but were conducted by PoliMI staff). INAF and PoliMI involved several SME for the procurements of components, systems and subsystems. For the P/L the mechanical parts were produced by Novomeccanica while the electronic boards were produced by CISTELAIER. For the S/C COTS components were provided by DHV, and GomSpace, in addition to SkyLabs. In all cases the contracts were managed by INAF and POLIMI. Acting as integrators gave us much more flexibility and the possibility of a very close link between the technological and the scientific teams. Of course, this also implies much lower costs than when the large industry or even SMEs act as integrators. This approach supported the design, integration and test of six complete S/C+P/L FM plus the P/L FM hosted by SpIRIT well. Scaling up to a more ambitious constellation (more complex and/or more numerous satellites to be developed in a shorter timescale) would probably require rethinking the organizational structure and project management, while maintaining the current flexibility and low costs.

## 6. Conclusions

HERMES-PF and SpIRIT were conceived at the end of the 2010s to show that high energy cosmic transients such as GRB and the electromagnetic counterparts of GWE can be routinely detected by miniatured hardware and localized with good accuracy using triangulation techniques. SpIRIT has validated the P/L design and showed that it is robust against the harsh space environment. HERMES-PF/SpIRIT will be the first relatively large constellation to monitor the high-energy sky, at least in the Western world. A similar constellation, GRID, has been recently launched by China [38,39], although born as an educational project. Several high-energy astrophysics CubeSats have been developed in Europe and the United States in the past years [40]. A particular mention must go to GRBAlpha, a 1U CubeSat launched in April 2020 and still working smoothly after more than three years in orbit [41,42,43]. GRBAlpha has detected so far 140 transients, including 83 GRB, so far. GRBAlpha has shown that miniaturized instrumentation hosted by CubeSats can routinely detected GRBs. So far, neither GRID nor Western CubeSats have produced a transient localization using the triangularization technique. This remaining challenge is the primary goal of the HERMES-PF/SpIRIT constellation.

**Acknowledgements**

This research has been possible thanks to the effort of the full HERMES-Pathfinder and SpIRIT teams. We acknowledge support from the European Union Horizon 2020 Research and Innovation Framework Programme under grant agreements n. 821896 HERMES Scientific Pathfinder and n. 871158 AHEAD2020, by ASI-INAF and ASI-POLIMI Accordi Attuativi "HERMES–Technologic Pathfinder" and by ASI-INAF Accordo Attuativo "HERMES Pathfinder–Operazioni e sfruttamento scientifico".